\documentclass[twocolumn,astrosymb]{aastex631}

\usepackage{amsmath}

\shorttitle{Carving the Edges of Rocky Planet Population}
\shortauthors{E. J. Lee \& J. E. Owen}

\begin{document}

\title{Carving the Edges of the Rocky Planet Population}

\author[0000-0002-1228-9820]{Eve J.~Lee}
\affiliation{Department of Astronomy \& Astrophysics, University of California, San Diego, La Jolla, CA 92093-0424, USA}
\affiliation{Department of Physics and Trottier Space Institute, McGill University, 3600 rue University, H3A 2T8 Montreal QC, Canada}
\affiliation{Trottier Institute for Research on Exoplanets (iREx), Universit\'e de Montr\'eal, Canada}

\author[0000-0002-4856-7837]{James E.~Owen}
\affiliation{Imperial Astrophysics, Department of Physics, Imperial College London, Prince Consort Road, London SW7 2AZ, UK}
\affiliation{Department of Earth, Planetary, and Space Sciences, University of California, Los Angeles, CA 90095, USA}

\begin{abstract}

Short-period planets provide ideal laboratories for testing star-planet interaction. Planets that are smaller than $\sim$2$R_\oplus$ are considered to be largely rocky either having been stripped of or never having acquired the gaseous envelope. Zooming in on these short-period rocky planet population, clear edges appear in the mass-period and radius-period space. Over $\sim$0.2--20 days and 0.09--1.42$M_\odot$, the maximum mass of the rocky planets stay below $\sim$10$M_\oplus$ with a hint of decrease towards $\lesssim$1 day, $\gtrsim$4 day, and $\lesssim 0.45 M_\odot$. In radius-period space, there is a relative deficit of $\lesssim$2$R_\oplus$ planets inside $\sim$1 day. We demonstrate how the edges in the mass-period space can be explained by a combination of tidal decay and photoevaporation whereas the rocky planet desert in the radius-period space is a signature of magnetic drag on the planet as it orbits within the stellar magnetic field. Currently observed catastrophically evaporating planets may have started their death spiral from $\sim$1 day with planets of mass up to $\sim$0.3$M_\oplus$ under the magnetic drag.
More discoveries and characterization of small planets around mid-late M and A stars would be welcome to better constrain the stellar parameters critical in shaping the edges of rocky planet population including their UV radiation history, tidal and magnetic properties.

\end{abstract}

\section{Introduction} \label{sec:intro}

Small planets abound in our Galaxy. With the detection of a clear gap that separates planets of size $\lesssim$ 1.6--2$R_\oplus$ and those of size $\sim$ 2--4$R_\oplus$ \citep{Fulton17,VanEylen18,Hsu19} in radius-period space, we now have statistical confirmation of two distinct populations.  The smaller group is generally considered to be rocky terrestrial planets, corroborating previous expectations from mass-radius relations \citep{Rogers15}. Under the mass loss origin of the radius gap, the population of the rocky planets may be considered as stripped cores of larger sub-Neptunes \citep[e.g.,][]{Valencia2010,Owen13,Owen17,Gupta2019,Rogers21}. Under the primordial origin of the radius gap, cores smaller than $\sim$1--2$M_\oplus$ cannot accrete sufficient nebular gas to emerge as sub-Neptunes, so they are born practically rocky \citep{Lee21,Lee22}. The two channels are not mutually exclusive: the primordial radius gap can be further sculpted by post-formation mass loss, which contributes particularly to the short period ($\lesssim$ 30--50 days) of rocky planet population \citep[e.g.,][]{Lee22}. Indeed \citet{Rogers21} required some fraction of the super-Earth population to be born with negligible initial hydrogen atmospheres to be consistent with the population level statistics. 

Although these short-period rocky planets may not retain their initial memory of formation, they are ideal laboratories for testing star-planet interaction, including the XUV stellar flux-driven photoevaporation \citep[e.g.,][]{Owen17,Jin18}, tidal decay and circularization \citep[e.g.,][]{Lee17,Petrovich19,Pu19}, and magnetic drag as the planet orbits through the stellar magnetic field \citep[e.g.,][]{Laine12,Lai12}. The latter two effects in particular can lead to the destruction of the planet as the planet reaches the Roche lobe limit, so we may expect their signature in terms of ``edges'' in the planet population. 

Recently, \citet{Parc24} suggested that there exists an upper limit to the mass of rocky planets at $\sim$10$M_\oplus$, irrespective of the mass of the host star over $\sim$0.2--30 days (see their Figure 2; see also our Figure \ref{fig:Mp_Per_Mstar}). Although this mass is close to the typically quoted critical core mass that triggers runaway gas accretion \citep[e.g.,][]{Mizuno80,Stevenson82,Pollack96}, runaway is likely not the cause of such a sharp edge to the planet mass distribution given how the critical core mass is expected to vary with the composition of the planet through opacity \citep[e.g.,][]{Ikoma00,Rafikov06,Lee14,Piso15,Savignac24} and we find plenty of sub-Neptunes (i.e., those that did {\it not} undergo runaway) with a mass exceeding $\sim$10$M_\oplus$ \citep[see][their Figure 2]{Parc24}. Instead, in this paper, we consider the interplay between photoevaporation and tidal decay in creating a sharp cutoff to the maximum mass of short-period rocky planets.

Yet another edge to the rocky planet population is the hint of a desert at $\lesssim$1--3 days and $\lesssim$1$R_\oplus$ in the radius-period space (see Figure \ref{fig:RvP-rocky-alfven}; see also \citealt{Dattilo23}, their Figure 6). This desert cannot be due to mass-loss as these planets are already rocky, and they are also not likely to be due to tidal decay, given the edge populations are smaller than Earth and therefore too small/light to drive sufficient tidal dissipation in the star to effect an orbital decay over $\sim$Gyr. Furthermore, this is unlikely to be a detection bias, as these small planets are detected at longer orbital periods.\footnote{However, one must consider that ultra-short-period planets can be hard to detect due to the finite cadence of the {\it Kepler} missions \citep{SanchisOjeda2014}.} Instead, we note that approximately half of these ultra-short-period sub-Earths are around low-mass stars with mass below $\sim$0.6$M_\odot$ which would be subject to a strong stellar magnetic field \citep[e.g.,][]{Shulyak19} and the planet at short orbital periods would be subject to orbital decay by magnetic drag \citep[e.g.,][]{Laine12,Lai12}, similar to the orbital decay of Echo 1 satellite through the Earth ionosphere \citep{Drell65} and the electrodynamic interaction between Io and Jupiter \citep{Goldreich69}.

This paper is organized as follows. In Section \ref{sec:sketch}, we outline the underlying theory deriving analytically the expected shape of the edges of short-period rocky planets in the mass-period and radius-period space. We verify our expectation with detailed numerical modeling in Section \ref{sec:simulations}. Summary and discussion are provided in Section \ref{sec:concl}.

\section{Theory} \label{sec:sketch}

We first begin by discussing the basic physics governing the star-planet interactions of close-in planets and present analytic arguments to derive the edges in mass-period and radius-period space. We then use our analytic boundaries to compare to the observed data, demonstrating that mass loss, tides, and magnetic interactions all likely play a role in sculpting the close-in rocky planet population. 

\subsection{Maximum core mass}
\label{ssec:theor-max-core}

\subsubsection{Photoevaporative mass loss}
\label{sssec:theor-photoevap}

We first discuss why having 10$M_\oplus$ as the maximum core mass of rocky planets that is weakly variant with orbital period and stellar mass inside orbital periods of $\sim$3--4 days is surprising in the context of photoevaporative mass loss alone. Following energy-limited approximation,
\begin{equation}
    \frac{GM_p \dot{M}_{\rm gas}}{R_p} = \frac{\eta}{4}L_{\rm XUV}\left(\frac{R_p}{a_p}\right)^2
    \label{eq:pevap}
\end{equation}
where $G$ is the gravitational constant, $M_p$ is the mass of the planet (roughly the core mass), $\dot{M}_{\rm gas}$ is the rate of envelope mass loss, $R_p$ is the radius of the planet, $\eta$ is the efficiency factor parametrizing the fraction of stellar irradiation used to liberate planetary envelope, $L_{\rm XUV}$ is the stellar XUV luminosity, and $a_p$ is the orbital distance of the planet. We do not account for the difference between the visible and XUV photospheres of the planet whose correction we consider subsumed in $\eta$.

Although $\eta$ is often approximated as a constant, in reality, it can vary with the irradiation flux and the gravitational potential of the planet in a complex manner \citep[e.g.,][]{Murray-Clay09,Owen12,Owen16} and can be expressed as a way to encompass the parameter space where the energy-limited approximation no longer holds \citep[e.g.,][]{Kubyshkina18,Caldiroli22}. Following \citet{Owen17}, their equation 31, we adopt the simplified expression
\begin{equation}
    \eta \sim 0.1\left(\frac{M_p}{10M_\oplus}\right)^{-1}
    \label{eq:eta}
\end{equation}
accounting for the difficulty of removing the envelope from a deeper potential well of a more massive planet. This functional form does not encode its full variation, but is appropriate in the vicinity of the radius-gap we are considering here (and explains the slope of the radius-gap, \citealt{VanEylen18}). Writing $\eta$ this way, we implicitly assume the planet mass to be core-dominated and the optical photosphere to be insignificantly dependent on the core mass \citep[e.g.,][]{Lopez14}.

The stellar XUV luminosity is expected to vary with both stellar mass and time, following
\begin{equation}
    \frac{L_{\rm XUV}}{L_\star} =
    \begin{cases}
        10^{-3.6} & t < t_{\rm sat} \\
        10^{-3.6} \left(\frac{t}{t_{\rm sat}}\right)^{-\alpha} & t \geq t_{\rm sat}
    \end{cases}
    \label{eq:Lxuv}
\end{equation}
where $t_{\rm sat}=$100 Myr is the saturation interval of high energy flux, and $L_\star$ is the bolometric luminosity \citep{Vilhu1987,Wright2011}. The power-law slope $\alpha$ ranges from 1.2 to 1.5 for X-ray luminosity \citep[e.g,][]{Ribas2005,Jackson2012,Claire2012} with a much shallower slope for EUV luminosity \citep[e.g.,][]{King21}, indicating that beyond 100 Myr, EUV dominates over X-ray. The full XUV luminosity therefore decreases with a shallower $\alpha = 0.86$ \citep[see][their Figure 1]{King21}, implying that the cumulative irradiated energy will be dominated at later times. In terms of the dependence on stellar mass, during the saturation phase, $L_X \propto L_\star \propto M_\star^3$ where the latter proportionality derives from energy transport by radiative diffusion with Thompson opacity, expected for main-sequence stars (we also verify with MIST stellar models; \citealt{Dotter16,Choi16}). Over time, the EUV spectrum becomes softer and $L_{\rm EUV} \propto L_X^{0.24}$ \citep{King21,Karalis24} in the soft wavelength (360-920 $\AA$). Combining everything, $L_{\rm XUV} \sim L_{\rm EUV} \propto M_\star^{0.72}$.

\begin{figure*}
    \centering
    \includegraphics[width=\textwidth]{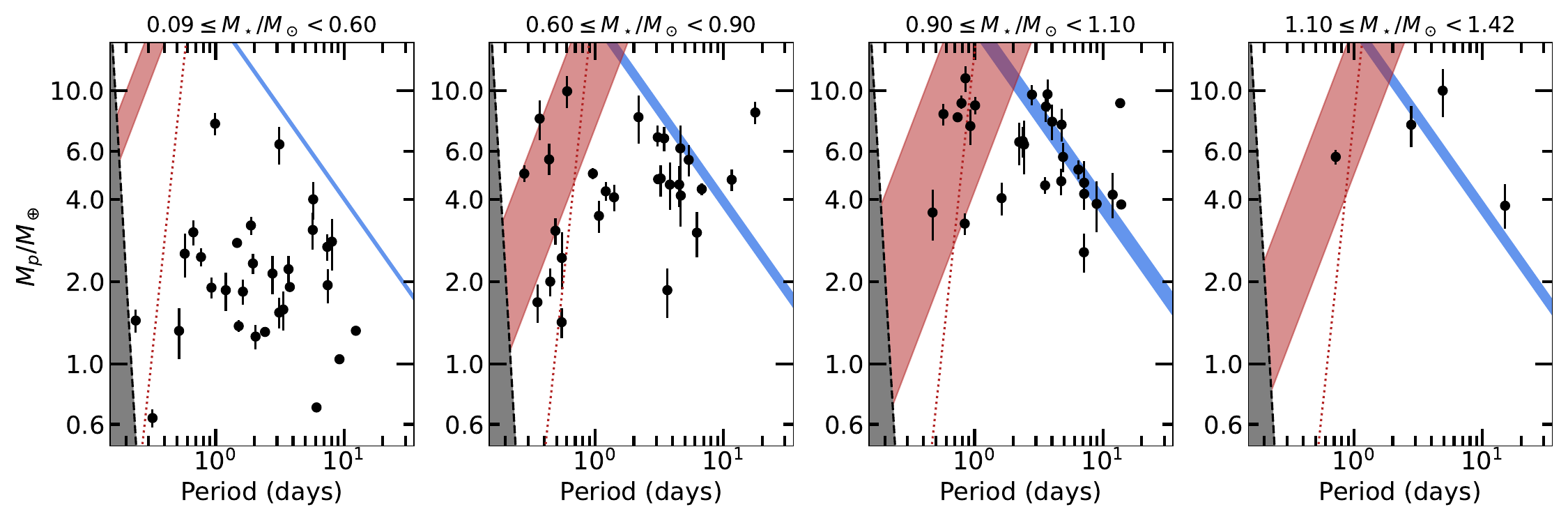}
    \caption{Planet mass vs.~orbital period for different host star mass, drawn from NASA Exoplanet Archive \citep{NASA-mass} using their default value. We only plot planets with radii $\leq 2R_\oplus$, period $<$20 days, and $\leq 25\%$ error in mass measurement. Blue region: maximum mass for complete photoevaporation with evaporation time spanning 9--14, 4--14, 2--14, and 2--7 Gyr for each stellar mass bin from lightest to heaviest (the choice of the age is from visual inspection of Figure 5 of \citealt{Petigura22}). Red region: characteristic maximum mass for tidal decay under dynamical tides for the same range of ages as the blue region. Red dotted: like the red region but for constant $Q'_\star=10^7$ for a fixed age of 5 Gyr, drawn for a reference. The grey zone is the forbidden zone due to Roche lobe overflow. The four stellar mass bins are chosen so that the number of planets within each bin is approximately uniform while ensuring the bin boundaries do not blur the stellar spectral types too much. The effect of choosing different bins is discussed in Appendix \ref{app:st-bin}.}
    \label{fig:Mp_Per_Mstar}
\end{figure*}

We now solve for $M_p$ below which its entire envelope will be evaporated over time $t$ by integrating $\dot{M}_{\rm gas}$ in equation \ref{eq:pevap} \citep{Mordasini2020}, with $\eta$ and $L_{\rm XUV}$ from equations \ref{eq:eta} and \ref{eq:Lxuv}, respectively, and $\alpha=0.86$:
\begin{equation}
    \frac{M_{\rm gas}(0)}{M_p}-\frac{M_{\rm lost,0}}{M_p}\left[\left(\frac{t}{t_{\rm sat}}\right)^{0.14}-0.86\right] \leq 0
\end{equation}
where $M_{\rm gas}(0)$ is the initial gas mass and
\begin{align}
    \frac{M_{\rm lost,0}}{M_p} &= 0.19 \left(\frac{\eta_0}{0.1}\right)\left(\frac{R_p}{3R_\oplus}\right)^3\left(\frac{M_p}{10M_\oplus}\right)^{-3} \nonumber \\
    &\times \left(\frac{P_p}{1\,{\rm day}}\right)^{4/3}\left(\frac{M_\star}{M_\odot}\right)^{0.05}\left(\frac{L_{\rm XUV,0}}{10^{-3.6}L_\odot}\right) \nonumber \\
    &\times \left(\frac{t_{\rm sat}}{100\,{\rm Myr}}\right)
\end{align}
with $M_\star$ being the mass of the host star, $P_p$ the orbital period of the planet, $\eta_0$ the numerical coefficient of equation \ref{eq:eta} and $L_{\rm XUV,0}$ the numerical coefficient of equation \ref{eq:Lxuv} anchored to $M_\star = M_\odot$, and $t_{\rm sat}=100$Myr. For $t = 1$Gyr,
\begin{align}
    M_{\rm p,loss} &\lesssim 15.0 M_\oplus \left(\frac{\eta_0}{0.1}\right)^{1/3}
    \left(\frac{M_{\rm gas}(0)/M_p}{0.03}\right)^{-1/3}
    \left(\frac{R_p}{3R_\oplus}\right) \nonumber \\
    &\times \left(\frac{P_p}{1\,{\rm day}}\right)^{-4/9}\left(\frac{M_\star}{M_\odot}\right)^{0.017}\left(\frac{L_{\rm XUV,0}}{10^{-3.6}L_\odot}\right)^{1/3}
\end{align}
where we set the planet's radius to a representative value to maximize the mass-loss timescale \citep{Owen17}. This maximum $M_{\rm p,loss}$ increases to $\sim$17.7$M_\oplus$ and $\sim$18.9$M_\oplus$ for $t=5$ and 10 Gyr, respectively. Under photoevaporation alone, we expect the maximum mass of the rocky cores to be larger than 10$M_\oplus$ and also increases toward shorter orbital periods, which we do not see within $\sim$1--3 days (Figure \ref{fig:Mp_Per_Mstar}). In fact, the original work predicting the radius gap showed cores of mass up to $\sim $15~M$_\oplus$ could be stripped at short periods \citep{Owen13}. 

\subsubsection{Orbital decay by tides}
\label{sssec:theor-tide-pl}

At ultra short orbital periods ($<$1 day), even rocky planets can undergo a significant tidal orbital decay \citep[e.g.,][]{Lee17} which can further sculpt the mass-period (and radius-period) relation.
Tides are raised in both the star and the planet, with the latter dominating when the planet's eccentricity is non-zero.
A planet can be driven to a non-zero eccentricity by, e.g., secular perturbation with neighboring planets. Most {\it Kepler} planets are expected to be in multi-planetary systems and their mutual interactions have been shown to be able to excite the innermost planet to moderate eccentricity, enough to drive tides within the planets, explaining ultra-short period planets \citep[e.g.,][]{Petrovich19,Pu19}. Nevertheless, we will show that the eccentricity tide raised on the planet likely does not play a major role in shaping the maximum mass of the rocky planet.

The tidal circularization time is
\begin{align}
    t_{\rm tide,p} &\sim 1.2\times 10^5\,{\rm yr} \left(\frac{Q_p/k_{L,p}}{100}\right)\left(\frac{R_p}{R_\oplus}\right)^{-5} \left(\frac{M_p}{M_\oplus}\right) \nonumber \\
    &\times \left(\frac{M_\star}{M_\odot}\right)^{2/3}\left(\frac{P_f}{1\,{\rm day}}\right)^4 \left(\frac{P_p}{10\,{\rm day}}\right)^{1/3}\left(\frac{F(0.9)}{F(e_p)}\right)
    \label{eq:t_tide_p}
\end{align}
where $Q_p$ is the planet's tidal quality factor, $k_{L,p}$ is the planet's Love number, $P_p$ is the initial period of the planet, $P_f=(1-e_p^2)^{3/2} P_p$ is the final circularized period of the planet, and 
\begin{equation}
    F(e) = 1 + \frac{31}{2}e^2 + \frac{255}{8}e^4 + \frac{185}{16}e^6 + \frac{25}{64}e^8
\end{equation}
from \citet{Hut81}.

We are looking for the maximum mass for survival so we take $P_f$ as the Roche lobe radius
\begin{align}
    P_R &= \frac{2\pi}{(GM_p)^{1/2}}q^{3/2}R_p^{3/2} \nonumber \\
    &\sim 0.22\,{\rm day} \left(\frac{q}{2.44}\right)^{3/2} \left(\frac{M_p}{M_\oplus}\right)^{-1/2} \left(\frac{R_p}{R_\oplus}\right)^{3/2}
    \label{eq:P_roche}
\end{align}
where $q$ is a numerical coefficient that depends on the structure of the planet, for which we adopt 2.44 following \citet{Rappaport13}.\footnote{Varying values are adopted in the literature from 2.7 \citep{Guillochon11} to 3.5 \citep{Matsakos16} which are based on the tidal disruption of gas giants. Adopting these large $q$ for rocky planets contradicts the existence of TOI-6255 b \citep{Dai24}, which is in our sample.} Plugging $P_f = P_R$ into equation \ref{eq:t_tide_p},
\begin{align}
    t_{\rm tide,p} &\sim 3.0\times 10^2\,{\rm yr} \left(\frac{Q_p/k_{L,p}}{100}\right) \left(\frac{q}{2.44}\right)^6 \left(\frac{R_p}{R_\oplus}\right) \left(\frac{M_p}{M_\oplus}\right) \nonumber \\
    &\times \left(\frac{M_\star}{M_\odot}\right)^{2/3} \left(\frac{P_p}{10\,{\rm day}}\right)^{1/3}\left(\frac{F(0.9)}{F(e_p)}\right).
    \label{eq:t_tide_rescl}
\end{align}
In reality, $P_p$ and $e_p$ are constrained by the expectation of tidal circularization, $P_R = (1-e^2)^{3/2}P_p$:
\begin{equation}
    1-e_p^2 \sim 0.08 \left(\frac{q}{2.44}\right) \left(\frac{M_p}{M_\oplus}\right)^{-1/3} \left(\frac{R_p}{R_\oplus}\right) \left(\frac{P_p}{10\,{\rm day}}\right)^{-2/3}
    \label{eq:eccen}
\end{equation}
corresponding to $e_p \sim 0.97$ for $M_p=10M_\oplus$ and $R_p \approx R_\oplus (M_p/M_\oplus)^{1/4} \sim 1.7R_\oplus$ for the chosen $M_p$. Numerical simulations of \citet{Petrovich19} show that such high eccentricities are possible for the innermost planet in a multiplanetary system through secular chaos, although this is demonstrated assuming the innermost planet has already started with $e \sim 0.5$. We find such a high initial eccentricity unlikely for the wide range of orbital period and stellar mass over which we find a uniform value of the observed maximum core mass. 

Furthermore, $t_{\rm tide,p}$ is way too short for $M_p=10M_\oplus$, suggesting that under the channel of tidal circularization, a 10$M_\oplus$ core would have quickly reached its Roche lobe limit and disintegrated. One possible way out is to consider tidal capture, in which the orbit shrinks faster than the periapse approaches the Roche radius under secular perturbation \citep[see][their equation 20]{Petrovich19}. However, the radius at which this tidal capture may occur $a_{\rm cap} \propto M_p^{-2/11}$ suggesting that the upper edge of the mass-period plot in Figure \ref{fig:Mp_Per_Mstar} would be drawn in the opposite direction (higher mass at shorter period) than what is observed. We therefore consider eccentricity tide an unlikely cause of the $\sim$10$M_\oplus$ edge---albeit still a viable process for contributing to the ultra-short period planet population---and proceed to consider tides raised on the star.

The orbital decay time under stellar tides is
\begin{align}
    t_{\rm tide,\star} &\sim 6.3\times 10^{9}\,{\rm yr} \left(\frac{Q'_\star}{10^7}\right) \left(\frac{M_\star}{M_\odot}\right)^{8/3} \left(\frac{R_\star}{R_\odot}\right)^{-5} \nonumber \\
    &\times \left(\frac{M_p}{10 M_\oplus}\right)^{-1} \left(\frac{P_p}{1\,{\rm day}}\right)^{13/3}
    \label{eq:t_inspiral}
\end{align}
where $Q'_\star$ is the tidal quality factor of the host star \citep[e.g.,][]{Goldreich66}. Compared to $t_{\rm tide,p}$, the orbital decay timescales under stellar tides are much closer to the typical ages of {\it Kepler} systems, especially those with planets inside orbital periods of $\sim$ 1 day \citep{Schmidt24}.

We search for $M_p$
so that $t_{\rm tide,\star} \gtrsim 5{\rm Gyr}$, where 5 Gyr is the median age of planetary systems shown in Figure \ref{fig:Mp_Per_Mstar} \citep[see also][their Figure 3]{Mcdonald19}:
\begin{equation}
    M_{\rm p,tide} \lesssim 12.6 M_\oplus \left(\frac{Q'_\star}{10^7}\right) \left(\frac{M_\star}{M_\odot}\right)^{-7/3} \left(\frac{P_p}{1\,{\rm day}}\right)^{13/3}.
    \label{eq:Mp-tide}
\end{equation}
Where $M_{\rm p,loss} > M_{\rm p, tide}$, all evaporated cores would undergo a significant tidal decay, which implies that inside
\begin{equation}
    P_p \lesssim 1.07\, {\rm day} \left(\frac{M_\star}{M_\odot}\right)^{0.49},
\end{equation}
the mass-period edge of rocky planets is sculpted by tides.

The steep dependence of $M_{\rm p,tide}$ on $P_p$ will manifest in the mass-period diagram as a sharp cliff with decreasing maximum $M_p$ at shorter orbital periods, dropping to $\sim$1.3$M_\oplus$ at just $\sim$0.6 days. We do not see this sharp cliff in the data. To recover the weak scaling between $M_p$ and $P_p$, we consider a tidal quality factor that is period-\textbf{dependent}.\footnote{The tidal quality factor is a parametrization of the fluid response to tidal forcing and is generally expected to be dependent on the forcing frequency---a constant Q' is often adopted mainly for simplicity \citep{Goldreich63}; see also review by \citet{Ogilvie14}.} Based on the empirical fitting to the hot Jupiters' orbital periods and mass, \citet{Penev18} report
\begin{equation}
    Q'_\star = 10^6 \left(\frac{P_p}{2\,{\rm days}}\right)^{-3.1}.
    \label{eq:Qstar-dyn}
\end{equation}
Plugging this relationship in to equation \ref{eq:Mp-tide}, the new maximum mass becomes
\begin{equation}
    M_{\rm p,tide} \sim 10.8\,M_\oplus \left(\frac{Q'_{\star,0}}{10^6}\right) \left(\frac{M_\star}{M_\odot}\right)^{-7/3} \left(\frac{P_p}{1\,{\rm day}}\right)^{1.23}
    \label{eq:Mp-tide-Q}
\end{equation}
where $Q'_{\star,0}$ is the numerical coefficient in equation \ref{eq:Qstar-dyn}. This much milder drop of $M_p$ with decreasing $P_p$ is more consistent with the data. Under a period-variant $Q'_\star$, tides sculpt the mass-period edge of rocky planets inside
\begin{equation}
    P_p < 1.34\,{\rm day} \left(\frac{M_\star}{M_\odot}\right)^{1.40}.
\end{equation}
While our derivation assumes that the tidal quality factor derived from the hot Jupiter sample applies identically to rocky planets, given that this $Q'_\star$ is that of the star and in both cases, $M_p/M_\star \ll 1$, we consider equation \ref{eq:Qstar-dyn} to also apply for the tidal decay of rocky planets.

Figure \ref{fig:Mp_Per_Mstar} summarizes our calculation, using the median $M_\star$ of the subset of data corresponding to each $M_\star$ bin. Our analytically derived maximum $M_p$-$P_p$ scaling compares remarkably well with the data within the variance of system age---the range for which we adopt from visual inspection of Figure 5 of \citet{Petigura22}---given the number of approximations we made.\footnote{Given the large range of system age and the relative sparseness of data, it is possible that a constant tidal quality factor $Q'_\star$ can also provide an adequate visual fit to the current data within the appropriate range of system age. We defer more detailed fitting to the data with a general parametrization of $Q'_\star \propto P_p^{n}$ to future work, although such an attempt would benefit from a large enough dataset of rocky planet masses to obtain reliable occurrence rates.} Our approach is further validated by the 
simulations in Section~\ref{sec:simulations}.

In the lowest $M_\star$ bin, $M_{\rm p,tide}$ (equation \ref{eq:Mp-tide-Q}) appears above the data, even at the estimated system age of 14 Gyr. This gap between theory and the current data could be due to difficulties in confirming ultra-short period planets and/or from the uncertain dependence of $Q'_\star$ on stellar mass---we would naively expect lower $Q'_\star$ around lower mass stars given their larger extent of the convective zone which would bring down $M_{\rm p,tide}$ (equation \ref{eq:Mp-tide-Q}) closer to the data in the lowest $M_\star$ bin. We further verify that all the planets in our sample lie away from the Roche limit (the one planet very close to the limit is TOI-6255 b around $M_\star=0.353 M_\odot$; \citealt{Dai24}).

There are a few planets that lie significantly above $M_{\rm p,loss}$. Given that $M_{\rm p, loss}$ is defined as the maximum mass of the core to have undergone complete stripping, planets that lie above it should be considered as those that still retain some amount of their volatile atmosphere. In the 0.60 $\leq M_\star/M_\odot \leq 0.9$ sample, these are HD 23472 b (8.32$^{+0.78}_{-0.79}M_\oplus$, 2.00$^{+0.11}_{-0.10}R_\oplus$; \citealt{Barros22}) and HD 136352 b (4.72$\pm 0.42 M_\oplus$, 1.664$\pm 0.043 R_\oplus$; \citealt{Delrez21}). More detailed interior modeling suggests that HD 23472 b may have a significant fraction of water and remaining gas consistent with the planet's placement above our $M_{\rm p,loss}$ \citep{Barros22}. In case of HD 136352 b, a dedicated mass loss modeling by \citet{Delrez21} finds that this planet likely lost the entirety of its atmosphere. While HD 136352 b is above our $M_{\rm p, loss}$, the deviation is within $\sim$2$\sigma$ which can likely be explained by the variation in the initial gas mass fraction (and therefore the initial planet radius), $\eta_0$ and $L_{\rm XUV}$.
In the 0.90 $\leq M_\star/M_\odot \leq 1.10$ sample, KOI-1599.02 (9.0$\pm 0.3 M_\oplus$, 1.9$\pm 0.2 R_\oplus$; \citealt{Panichi19}) is significantly above $M_{\rm p, loss}$.\footnote{Kepler-36 b also lies above this line at a level that is statistically significant but a dedicated evaporation models, with tabulated rather than scaled efficiencies, can identify this planet as an evaporated core \citep[e.g.,][]{Owen16-Kepler36}.} 
The mass of this planet is constrained by transit timing variation through a re-analysis of {\it Kepler} light curves \citep{Panichi19} and even the lower mass solution $\sim$6.8$M_\oplus$ would place this planet too high above our $M_{\rm p, loss}$. Given its relatively large size, KOI-1599.02 could also have retained some of its atmosphere, just like HD 23472 b. In the 1.10 $\leq M_\star/M_\odot \leq 1.41$ sample, Kepler-107 c (10.0$\pm 2.0 M_\oplus$; 1.597$\pm 0.026 R_\oplus$; \citealt{Bonomo23}) is more than 2-$\sigma$ away from our theoretical $M_{\rm p, loss}$. Compared to its inner neighbor planet b, Kepler-107 c is reported to be significantly denser suggesting its property is more likely sculpted by giant impact rather than photoevaporative mass loss \citep{Bonomo19}.

\subsection{Rocky Planet Desert}

We now consider an edge to the rocky planet population that appears in the radius-period space. Planet occurrence rate studies report a deficit of sub-Earths inside $\sim$3 days \citep[see, e.g.,][their Figure 6]{Dattilo23}, whose short periods assure this feature to be physical rather than a result of detection bias. Here, we describe how such an edge can be carved out by the magnetic interaction between the star and the planet.

\subsubsection{Magnetic drag}
\label{ssec:th-mag-drag}

The relative motion of the planet with respect to the stellar magnetosphere induces an electric field, current, and potential difference across the planet while perturbing the stellar magnetic field lines, driving Alfv\'{e}n waves \citep[e.g.][]{Strugarek2018}. These waves act to transmit information about the induced electric field between the planet and the star and can establish a ``closed circuit'' so long as the wave propagation time is shorter than the time it takes for the field line to slip past the planet \citep[e.g.,][]{Goldreich69,Laine12,Strugarek2014}. 
This causally connected star-planet magnetic interaction is ensured as long as the relative speed between the planet and the stellar magnetosphere is sub-Alfv\'enic, which we will show to be true. The planets we consider here have orbital periods $\lesssim$1--3 days, which are shorter than the typical spin periods of the stellar magnetosphere. We therefore expect magnetic torques to cause an orbital decay of the planets.\footnote{At these short orbital periods, the stellar winds are also expected to be sub-Alfv\'enic further ensuring the magnetic causal connection between the star and the planet \citep{Lanza2020}.}
In this picture, the loss of orbital energy is mediated by the Joule heating of the induced current whose radiated power is, in cgs units (c.f. \citealt{Laine12} for SI unit),

\begin{equation}
    {\cal P} = \frac{16 R^2_p (\Omega_p - \Omega_\star)^2 a^2_p B^2(a_p) \Sigma s}{c^2}
   \label{eq:P_LL}
 \end{equation}
where $\Omega_p \equiv \sqrt{GM_\star/a^3_p}$ is the planet's orbital frequency, $\Omega_\star$ is the stellar spin frequency, $B(a_p)$ is the local strength of the star's magnetic field at the planet's separation, $\Sigma$ is the spatial integral of the local Pedersen conductivity of the stellar surface, and $s \equiv \sqrt{1-R_\star/a_p}$ accounts for spherical geometry so that the resistance (not to be confused with resistivity which is independent of the geometry of the problem) is written as:
 \begin{equation}
     {\cal R} = \frac{1}{2\Sigma s}.
 \end{equation}

In writing ${\cal P}$ as the total Joule heating of the circuit, we have implicitly assumed ${\cal R}$ of the planet to be negligible compared to that of the star. The actual value of resistance is uncertain and depends on the conductivity profile of both the stellar atmosphere and the interior of the rocky planet (which may not actually be rocky). We will now re-write ${\cal P}$ in a way that avoids the direct calculation of ${\cal R}$.

From Ohm's law, $I = \Sigma U$ where $I$ is the induced current and $U$ is the induced potential difference across the surface of the planet. This potential difference can be written as
\begin{equation}
    U \sim -E_{\rm ind} R_p \sim v_k B(a_p) R_p
\end{equation}
where the second equality follows from the motional field  $\vec{E} = \vec{v}\times\vec{B}/c$ and $\vec{v}$ is the velocity of the conductor (i.e., the planet). From Faraday's law,
\begin{equation}
    E \sim \frac{B}{c} \left(\frac{L}{t}\right)
    \label{eq:faraday}
\end{equation}
where $L$ is a characteristic length scale and $t$ is a characteristic time scale. From Amp\`ere's law,
\begin{equation}
    B \sim \frac{4\pi}{c} J L
    \label{eq:ampere}
\end{equation}
where $J$ is the current destiny. Combining everything,
\begin{align}
    \left(\frac{L}{t}\right)^{-1} v_k B(a_p) &\sim \frac{4\pi}{c^2} \frac{I}{L} \nonumber \\
    \frac{v_k}{v_A} B(a_p) &\sim \frac{4\pi}{c^2} \frac{I}{R_p} \nonumber \\
    I &\sim \frac{c^2}{4\pi} R_p B(a_p) \frac{v_k}{v_A}.
\end{align}
We take $L/t$ as the Alfv\'en velocity $v_A$ on the left hand side because this current is the current required to maintain the same \textbf{$E \sim v_k B(a_p)/c$} as the planet's orbit continuously generates Alfv\'en disturbance which propagates at speed $v_A$. We take $L \sim R_p$ on the right hand side as this $E$ is across the surface of the planet. We can now solve for our estimate of $\Sigma$:
\begin{align}
    \Sigma &= \frac{I}{U} \nonumber \\
    &\sim \frac{c^2}{4\pi v_A}. 
    \label{eq:sigma-resist}
\end{align}
Plugging this into ${\cal P}$,
\begin{equation}
    {\cal P} \sim \frac{32}{3\pi} \pi R_p^2 v_A \left(\frac{v_k}{v_A}\right)^2 \frac{B^2(a_p)}{8\pi} s,
    \label{eq:P_mag_drag_scl}
\end{equation}
where $\Omega_p \gg \Omega_\star$ limit is taken as appropriate for planets inside $\sim$1 day around stars of typical spin period $\sim$10 days. 

Assuming this radiation purely removes orbital energy, the orbital distance evolves following
\begin{equation}
    \frac{GM_\star M_p}{2a_p^2}\dot{a}_p = -{\cal P}.
    \label{eq:orbevol}
\end{equation}
We consider the stellar field to be dipolar and use
\begin{equation}
    B(a_p) = B_\star \left(\frac{R_\star}{a_p}\right)^3.
\end{equation}
The local Alfv\'en velocity $v_A \equiv B(a_p)/\sqrt{4\pi\rho_i}$, where $\rho_i$ is the density of the ionized medium, which we express using the properties of stellar wind:
\begin{equation}
    \rho_i = \frac{\dot{M}_w}{4\pi a_p^2 v_w}
\end{equation}
where $\dot{M}_w$ is the stellar mass loss rate by wind and $v_w$ is the speed of the stellar wind. As our fiducial values, we adopt $\dot{M}_w = 10^{-12} M_\odot\,{\rm yr}^{-1}$ and $v_w = 100\,{\rm km\,s^{-1}}$ from \citet{Johnstone15-1} and \citet{Johnstone15-2}. We then obtain
\begin{align}
    {\cal P} &\sim 9.03\times 10^{23}\,{\rm erg\,s^{-1}} \left(1-\frac{R_\star}{a_p}\right)^{1/2}\left(\frac{B_\star}{100\,{\rm G}}\right)\left(\frac{R_\star}{R_\odot}\right)^3 \nonumber \\
    &\times \left(\frac{M_\star}{M_\odot}\right)^{-2/3}\left(\frac{P_p}{1\,{\rm day}}\right)^{-10/3}\left(\frac{R_p}{R_\oplus}\right)^2 \nonumber \\
    &\times \left(\frac{\dot{M}_w}{10^{-12}M_\odot\,{\rm yr}^{-1}}\right)^{1/2}\left(\frac{v_w}{100\,{\rm km\,s^{-1}}}\right)^{-1/2}.
    \label{eq:Psat}
\end{align}
With our choice of parameters, $v_k/v_A \sim 0.13$ at 1 day and decreases as the planet spirals in, scaling with $v_k/v_A \propto P_p$ so the star-planet magnetic interaction remains sub-Alfv\'{e}nic for the entirety of orbital decay, which is a requirement to sustain a ``closed circuit'', validating our assumption. We highlight how this is in contrast to \citet{Lai12} who derived a limit where the circuit is no longer closed due to order unity twisting of the magnetic field. The degree of this twist is on the order of $v_k/v_A$ which is $<1$ in our case so our treatment of the magnetic drag is self-consistent. In other words, this sub-Alfv\'{e}nic criterion ensures that the total ${\cal R}$ in the circuit is high enough to have minimal twist in the flux tube (compare equation \ref{eq:sigma-resist} with equation 9 of \citealt{Lai12}).

So far, we assumed $B_\star$ to be temporally constant. However, like XUV luminosity, stellar magnetic field strength is expected to decline with time after the saturation phase \citep[e.g.,][]{Folsom16}. Observationally, this decline is seen most readily with respect to the stellar spin period with weaker magnetic fields (or activity signatures) for slower rotators \citep[e.g.,][]{Pizzolato03,Wright18,Shulyak19}. From semi-empirical models of stellar spin evolution \citep{Gallet13,Ardestani17}, beyond the saturation phase, $B_\star \propto Ro^{-1.2}$ where $Ro \equiv 2\pi / \Omega_\star \tau_c$ is the Rossby number and $\tau_c$ is the convective overturn timescale evaluated at the bottom of the convective zone. On the main sequence, stellar spin down is expected to follow the Skumanich law $\Omega_\star \propto t^{1/2}$ \citep{Skumanich72} so, altogether, $B_\star \propto t^{-0.6}$ (assuming $\tau_c$ is constant), which is close to the measured $B_\star \propto t^{-0.655 \pm 0.045}$ \citep{Vidotto14}. 
Writing $B_\star = B_{\star,i} (t/t_{\rm sat})^{-0.6}$ with $t_{\rm sat}=$100 Myr, and
\begin{equation}
    {\cal P} =
    \begin{cases}
         {\cal P}_{\rm sat} & t < t_{\rm sat} \\
        {\cal P}_{\rm sat} \left(\frac{t}{t_{\rm sat}}\right)^{-0.6} & t \geq t_{\rm sat},
    \end{cases}
    \label{eq:P_tvar}
\end{equation}
where ${\cal P}_{\rm sat}$ is equation \ref{eq:Psat} with $B_{\star,i}$ substituted in lieu of $B_\star$.

\begin{figure}
    \centering
    \gridline{\fig{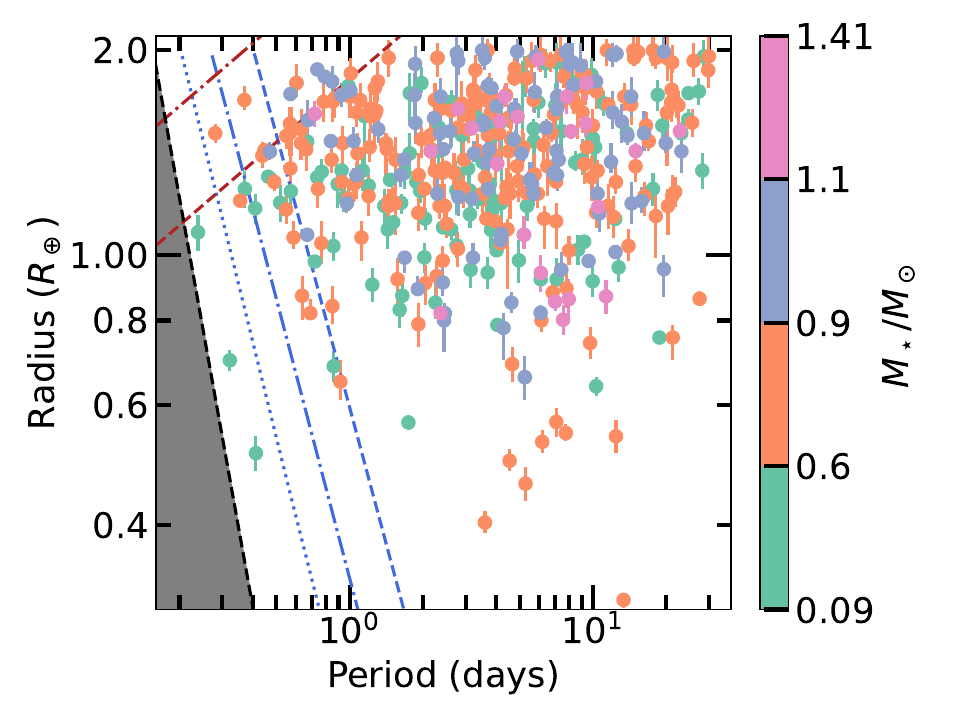}{0.5\textwidth}{}}
    \vspace{-1cm}
    \gridline{\fig{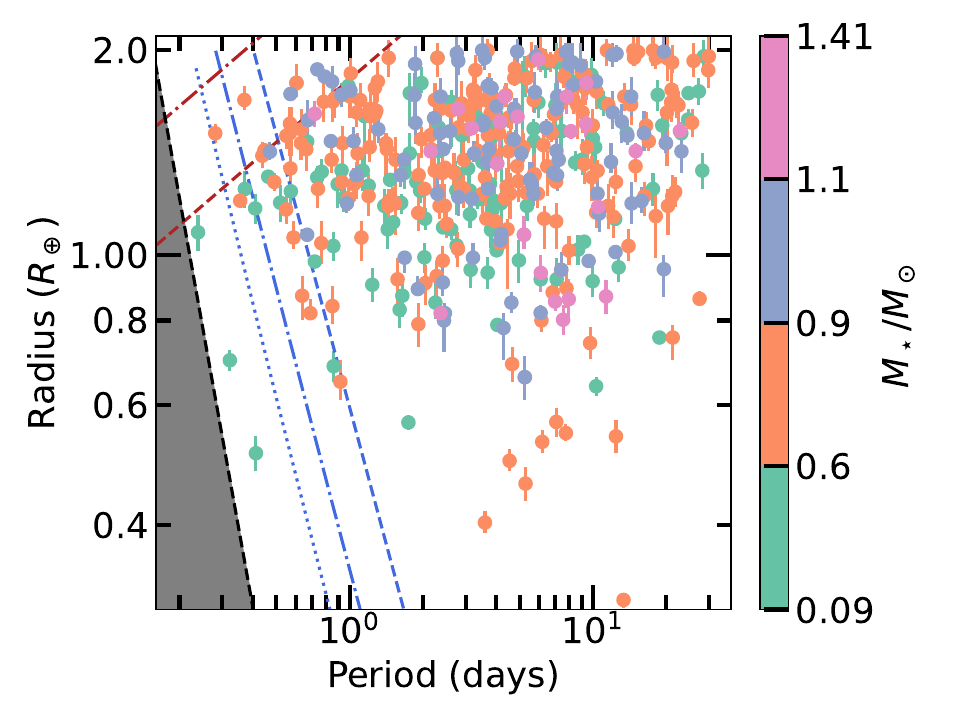}{0.5\textwidth}{}}
    \vspace{-0.8cm}
    \caption{Radius vs.~period drawn from NASA Exoplanet Archive \citep{NASA-rad} using their default value. We only plot planets with $\leq$8\% error in radius measurement. Top: the blue lines draw the minimum survival radius for time-varying $B_\star$ with $B_{\star,i} =$ (100, 30, 10) G for dashed, dot-dashed, and dotted lines, respectively. All other parameters are set to fiducial values. The red dashed and dot-dashed lines show the radius of the planet corresponding to the maximum mass that survives against tidal decay (equation \ref{eq:Mp-tide-Q}) for 5 Gyr around 1$M_\odot$ and 0.5$M_\odot$ star, respectively. The grey zone is the forbidden region inside which a rocky planet would be tidally disrupted. Bottom: same as the top panel but now the blue lines correspond to varying stellar mass $M_\star = $(1, 0.5, 0.3) $M_\odot$ for dashed, dot-dashed, and dotted lines, respectively, for $B_{\star,i} = 100 $G.}
    \label{fig:RvP-rocky-alfven}
\end{figure}

Integrating equation \ref{eq:orbevol},
\begin{align}
    &\int^{a_p(t)/a_0}_{a_p(0)/a_0} \left(1-\frac{R_\star}{a_p}\right)^{-1/2}\left(\frac{a_p}{a_0}\right)^3 d\left(\frac{a}{a_0}\right) \nonumber \\
    &=-\frac{2a_0}{GM_\star M_p}{\cal P}_{\rm sat} t_{\rm sat} \times \left[-1.5+2.5\left(\frac{t}{t_{\rm sat}}\right)^{0.4}\right],
    \label{eq:integral-alfven}
\end{align}
where we take $a_0=0.0196$ au (1 day around a solar mass star). In the above equation, we assumed that $\dot{M}_\star$ and $v_{\rm w}$ are constant in time. Stellar wind models report time-varying $\dot{M}_\star$ between $\propto t^{-0.62}$ and $\propto t^{-1.23}$ \citep{Suzuki13,See14,Johnstone15-2} depending on the assumed evolution of stellar spin, magnetic topology, and wind temperature. Given the unresolved degeneracy between $\rho_i$ and $v_w$ when only the time-variance of $\dot{M}_w$ is known (which itself is uncertain), we ignore this complexity in our simple approach.

A rocky planet is expected to survive if $a_p(t)$ is greater than the Roche limit (equation \ref{eq:P_roche}) so we solve for this inner $R_p$-$P_p$ edge by setting $a_p(t)$ to the Roche limit, $t$ as the system age (fiducial value of 5 Gyr) and find $R_p$ such that the above equation holds. Our solution is numerical since the integral on the left-hand side of equation \ref{eq:integral-alfven} is a complicated function (see equation \ref{eq:integral-analytic} for the full expression) and it is not possible to derive a simple scaling relationship. Fitting a power-law relation to the numerically derived $R_p$-$P_p$ edge for $B_{\star,i}$=100 G, $M_\star=M_\odot$, and $t=5$ Gyr (with $M_\star \propto R_\star$ as expected for main sequence stars and $M_p \propto R_p^4$ as expected for rocky bodies; \citealt{Valencia06}),
\begin{equation}
    R_{\rm p,mag} \sim 0.60 R_\oplus \left(\frac{P_p}{1\,{\rm day}}\right)^{-1.34}.
\end{equation}

We note that a simple scaling relationship between $R_p$ and $P_p$ can be obtained for survival against magnetic drag in the limit of $R_\star \ll a_p$. Inside orbital periods of $\sim$1 day that we study here, however, such an assumption becomes inaccurate, necessitating the numerical calculation as described above.

Figure \ref{fig:RvP-rocky-alfven} demonstrates another remarkable agreement between our semi-analytically derived $R_p$-$P_p$ edge and the data within the variance of $B_{\star,i}$ and the stellar mass. 
There are three planets that lie below our theoretical edge: TOI-6255 b, which we have already discussed, KOI-4777.01 (0.51$\pm 0.03 R_\oplus$, 0.41 days; \citealt{Canas22}), and GJ 367 b (0.699$\pm 0.024 R_\oplus$, 0.32 days; \citealt{Goffo23}). All three planets are around low-mass stars, which are generally expected to feature stronger magnetic fields with the dipole component dominating compared to high-mass stars \citep[e.g., see][their Figure 15]{Kochukhov21}. However, the reported spin periods of these latter two planet hosts are particularly slow (44 days and 51.3 days for KOI-4777 and GJ 367, respectively) which could imply a low $B_{\star,i}$. Although we were unable to find the spin period of TOI 6255 in the literature, it would be a good target to measure the stellar spin to verify whether TOI 6255 is also a slow rotator.
As indicated in Figure \ref{fig:RvP-rocky-alfven}, adopting lower $M_\star$ and $B_{\star,i}$ will both contribute to drawing the $R_p$-$P_p$ edge due to magnetic drag closer to the Roche limit and bring the analytic relation closer to the position of these three planets in the radius-period space.

\begin{figure*}
    \centering
    \includegraphics[width=\textwidth]{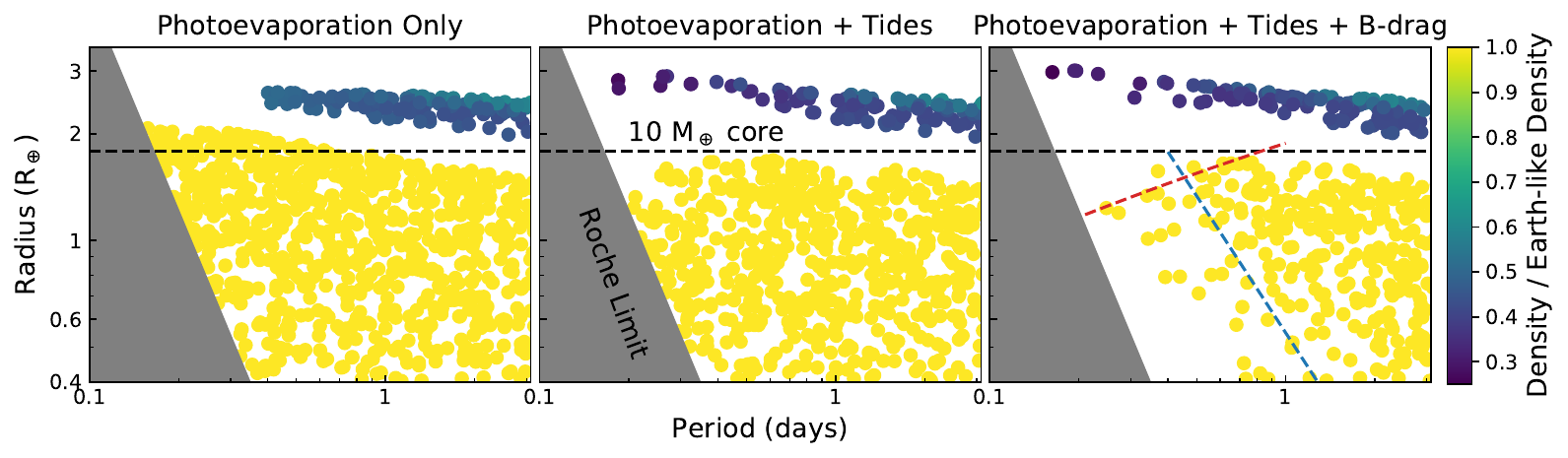}
    \caption{The distribution of planetary radii after 5 Gyrs of evolution from our numerical simulations for a solar mass star. The left panel includes the influence of photoevaporation only, demonstrating photoevaporation can create super-Earths with masses $> 10$~M$_\oplus$ at periods $\lesssim 1$~day. The combination of photoevaporation and tidal decay (middle panel) limits the maximum super-Earth mass to $\sim10$~M$_\oplus$. Finally, the combination of photoevaporation, tidal decay and magnetic drag (right panel) sculpts in close-in super-Earth population, carving demographic features similar to the observed population. The addition of tidal decay does populate the evaporation desert with the odd planet that has retained its H-dominated atmosphere; these planets have been caught spiraling into the host star. The grey zone is the forbidden region inside which a rocky planet would be tidally disrupted.  The dashed red and dashed blue lines in the right panel are our tidal and magnetic decay boundaries, respectively (same as those in the top panel of Figure~\ref{fig:RvP-rocky-alfven}). Note the significant drop in planet density inside these boundaries, planets found at shorter periods than these lines are undergoing significant orbital decay on timescales $\lesssim 1$~Gyr. }
    \label{fig:simulations}
\end{figure*}

\section{Simulations}\label{sec:simulations}
In order to verify our analytic expectations, we simulate the evolution of the close-in \textbf{exoplanet} population under the influence of photoevaporation, dynamic tides, and magnetic drag, co-evolving planetary interiors and orbits. For the planetary interior evolution calculations, we use the numerical model of \citet{Owen17}, but with updated mass-loss efficiencies from tabulated photoevaporation models implemented by \citet{Rogers2024}. 

To evolve planetary orbits, we solve the ordinary differential equation (ODE) ${\rm d}a/{\rm d}t$ including both dynamic tides (equations \ref{eq:t_inspiral} and \ref{eq:Qstar-dyn}) and magnetic drag (equations \ref{eq:orbevol} and \ref{eq:P_tvar}). The ODE for the orbital evolution is solved using Euler time-stepping, where the timestep is limited so that the planet's semi-major axis can only change by a maximum of 1\% per timestep. If required, we use subcycling of the orbital decay ODE between interior evolution time steps. For orbital evolution, we adopt the limit $\Omega_p\gg\Omega_*$ such that $\Omega_p\approx\Omega_p-\Omega_*$ for the entire evolution, which is appropriate, since the orbital evolution of those planets we are interested in here have periods $\lesssim 1~$ day, much faster than a typical slow rotating star with rotation periods $\gtrsim 10~$days. 

Our goal with the simulations here is to verify the analytic results, rather than to explore the parameter space to ``fit'' the exoplanet distribution (such an effort is extremely computationally expensive; see, e.g., \citealt{Rogers21}). Thus, we choose to simulate a planet population around a solar mass star, with initially uniform distribution in both orbital period and core-radius, between the Roche limit and 4~days and between 0.4 and 2.1~R$_\oplus$ ($\sim 20$~M$_\oplus$), respectively. Although there is clear evidence that the occurrence rate of super-Earths and sub-Neptunes increases with longer periods out to $\sim$10 days \citep[e.g.][]{Fressin13,Petigura18,Wilson22}, and that the core-radius/mass distribution is not uniform \citep{Rogers21,Lee21}, using a uniform distribution allows us to clearly see the edges carved in the distribution by the different mechanisms. Furthermore, we adopt a fixed initial H/He mass fraction of 3\% and the fiducial parameters for the strength of the dynamic tides and magnetic drag (see Section \ref{sec:sketch}). We verify that our simulation results are independent of the choice of initial gas mass fraction provided it lies between $\sim$1--100\%, consistent with the expectations of the photoevaporation theory \citep{Owen17,Owen2019,Mordasini2020}.

Figure~\ref{fig:simulations} corroborates the analytic derivations in Section \ref{sec:sketch}, demonstrating its consistency with the trends seen in the exoplanet population. As expected, photoevaporation alone readily creates massive super-Earths ($>10M_\oplus$), provided they exist well inside 1 day orbit initially. However, with the introduction of tides, those planets with cores $\gtrsim 10$~M$_\oplus$ 
tidally decay and are destroyed at the Roche limit. Finally, once magnetic drag is included, planets $\lesssim 1$~R$_\oplus$ inside $\sim$1 day orbit undergo a significant orbital decay, drastically reducing their abundance. We also verify with our simulations the weak scaling of the maximum super-Earth mass on stellar age and stellar mass found from the analytic calculations. 

Our calculations here focus on reproducing the ``edges'' of the population. To reproduce the overall shape of the mass-radius-period distribution, incorporating mass-loss and orbital evolution over a large swath of the parameter space with physically-derived or empirical prior would be needed, including the distribution of the stellar spin \citep[e.g.,][]{Lee17,Kubyshkina2019}.

\begin{figure*}
    \centering
    \includegraphics[width=\linewidth]{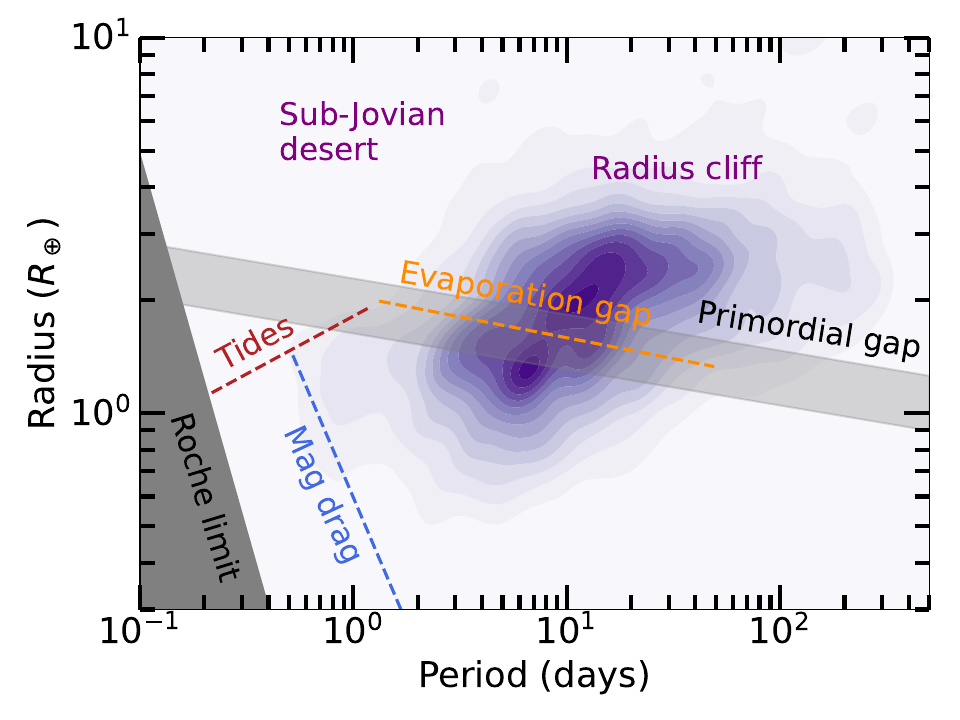}
    \caption{A grand summary highlighting all the physical processes that sculpt the small planet population in the radius-period space. Plotted in contour is the Gaussian Kernel density estimation of {\it Kepler} candidates from Data Release 25 \citep{Thompson18}. We do not correct for detection biases as these contours are for illustration purpose. Rocky planets are expected to be found in a region bounded by the evaporation/primordial gap (orange dashed line / light grey region), tides (red dashed line), and magnetic drag (blue dashed line). All the boundaries drawn in this figure assume system age of 5 Gyr, $M_\star=M_\odot$, and $B_{\star,i}=$100 G. Varying these stellar parameters can widen or shrink the allowable region of surviving rocky planets.}
    \label{fig:grand-summary}
\end{figure*}

\section{Discussion and Conclusion} \label{sec:concl}

In this paper, we demonstrated how the radiative, tidal, and magnetic interactions between the host star and an orbiting planet can sculpt the three observed edges in the short-period, rocky planet population. In spite of our simplifying assumptions, our analytic derivations, verified by numerical simulations, show a remarkable agreement with the data, within the variance of uncertain parameters such as the system age and the initial stellar magnetic field intensity.

In Figure \ref{fig:grand-summary}, we illustrate all the physical processes that sculpt the small planet population in the radius-period space. In addition to what we discussed in this paper, we also indicate the primordial gap that separates practically rocky planets from gas-enveloped planets at birth \citep{Lee21,Lee22} whose features can be tested at long orbital periods $\gtrsim$100 days where the initial memory of formation can still be retained against mass loss processes owing to lower stellar irradiation there. The three limits derived from first principles---evpaoration/primordial gap, tidal decay, and magnetic drag---outline the boundary of a region inside which we expect to find surviving rocky planets. For clarity, we choose our fiducial value (5 Gyr, 1$M_\odot$, $B_{\star,i}=100$G) for Figure \ref{fig:grand-summary}. In reality, the variance in system and stellar parameters can shift around these boundaries enlarging or shrinking the area of the parameter space in which we expect the rocky planets to reside.

In the context of the overall population expanding to larger sub-Neptunes and Saturns, we also find the sub-Jovian desert (the lack of Saturn-sized objects at short orbital periods) and the radius cliff (the steep drop off in the radius distribution beyond $\sim$4$R_\oplus$). While Jupiters are massive enough to withstand photoevaporative mass loss even at these short orbital periods \citep{Murray-Clay09}, Saturns are more vulnerable. The lower edge of the sub-Jovian desert (i.e. hot Saturns and Neptunes) has been explained as being carved out by photoevaporation whereas the upper edge (i.e., hot Jupiters) by high-eccentricity migration followed by tidal orbital decay (\citealt{Owen18}; see also \citealt{Matsakos16}). Similarly, the radius cliff can also be explained by photoevaporation in combination with an underlying core mass distribution that falls off beyond $\sim$10$M_\oplus$ (\citealt{Hallatt22}; see also \citealt{Lee19}). \citet{Dattilo24} report how using the framework of \citet{Rogers21}, the observed morphology of the radius cliff cannot be fully explained by photoevaporation. 
Deriving fine structures in demographic patterns in the radius-period space, however, is sensitively determined by uncertain input parameters including the underlying core mass distribution. In fact, photoevaporation models are often fit against data to derive the underlying core mass distribution. Independently solving for the core mass distribution (whether by direct mass measurements of small planets or by ab initio core coagulation simulations) would be welcome to truly test the ability of evaporation to explain the radius cliff, as the disagreement found by \citet{Dattilo24} could be pointing to more complexity in the core-mass function than that assumed in \citet{Rogers21} model (e.g. a core-mass function that varies with orbital period); see also \citet{Rogers2023}.

\begin{figure}
    \centering
    \includegraphics[width=0.5\textwidth]{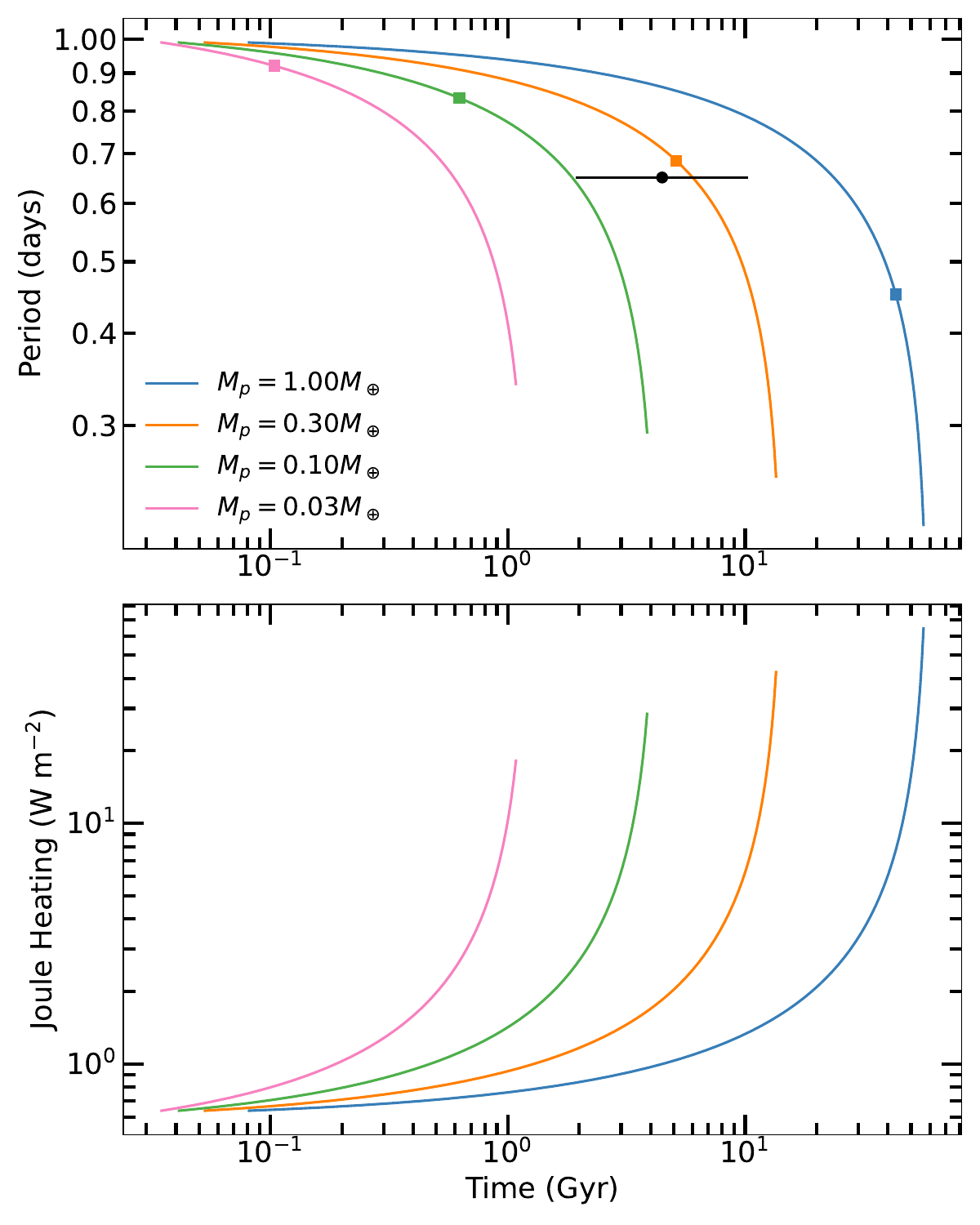}
    \caption{Top: orbital evolution of rocky planets under magnetic drag. Each curves terminate when the Roche radius is reached. Marked in squares are the point beyond which the Joule heating exceeds the planet's gravitational binding energy (see text for detail). The black circle indicates the orbital period and the age of the disintegrating planet Kepler-1520 b \citep{Rappaport12,Morton16}. Bottom: the Joule heat flux deposited into the planet along the orbital period evolution.}
    \label{fig:magdrag-decay}
\end{figure}

\subsection{Magnetic Drag, Disintegrating Planets,\\ and Lava Planets}

Figures \ref{fig:RvP-rocky-alfven} and \ref{fig:grand-summary} show that inside 1 day, rocky planets smaller than $\sim$1$R_\oplus$ can undergo a significant orbital decay by magnetic interaction rather than tidal interaction with the host star. The dominance of magnetism over tides contrasts with hot Jupiters whose tidal orbital decay dominates over magnetic drag by 4--6 orders of magnitude \citep{Lai12}. The switch from tide to magnetic drag from gas giants to rocky planets arises from the extreme sensitivity of the tidal torque on planet mass $\propto M_p^2$, which translates into $R_p^8$ for $M_p \propto R_p^4$, while the magnetic torque scales only as $R_p^2$. 

The efficiency at which the magnetic drag can bring sub-Earth planets into ultra-short orbital periods suggests that it may be a dominant mechanism to form disintegrating planets such as Kepler-1520 b \citep{Rappaport12}, KOI-2700 b \citep{Rappaport14}, and K2-22 b \citep{Sanchis-Ojeda15}. The shape and the variability of their transit depth can be explained by a cometary tail produced by vaporization given their extreme surface temperatures and proximity to the host star \citep[e.g.,][]{Perez-Becker13}. The vaporized outflow cools and condenses into dust grains that obscures the star \citep{Booth23,Campos24} and the heating-cooling cycle can give rise to chaotic bursts of winds \citep{Bromley23}. While the nature of these dying planets has been investigated in detail, how exactly the planets got to such short orbital periods received less attention. 

The fact that these disintegrating planets are losing their material necessarily implies their low mass ($\lesssim$0.1$M_\oplus$). Based on the evaporation time of Kepler-1520 b, \citet{Perez-Becker13} estimate that for every disintegrating planet, there should be $\sim$10-100 planets no larger than $\sim$0.4$R_\oplus$ at sub-day orbital periods. We see no such planets in the data and this paucity is likely physical in origin as such small planets cannot withstand magnetic drag down to the Roche limit within $\sim$5 Gyr (see Figure \ref{fig:RvP-rocky-alfven}).
Instead, we consider the possibility that the progenitor planet was even larger than Mercury (the maximal size required by catastrophic evaporation) and began its death spiral through the Joule heating and the magnetic drag.

Figure \ref{fig:magdrag-decay} illustrates the orbital evolution of rocky planets under the magnetic drag around 1$M_\odot$ star of $B_{\star,i} = $100 G. In the closed circuit established by the planet's orbit through the stellar magnetic field, both the stellar surface and the planet act as resistors. Using the conductivity of the present-day Earth mantle, \citet{Laine12} estimate the resistance of the planet to be roughly $4\times 10^{-3}$ that of the star.\footnote{The actual numerical value of ${\cal R}$ calculated by \citet{Laine12} is small enough to generate a significant twist in the flux tube breaking the field lines. At this point, the magnetic drag will be through radiation by Alfv\'en waves through open field lines with identical scaling relation as equation \ref{eq:P_mag_drag_scl}, although the numerical coefficient may be different.} We then solve for the time at which $4 \times 10^{-3}$ of the dissipated orbital energy exceeds the gravitational binding energy of the planet (marked with squares in Figure \ref{fig:magdrag-decay}). For the wide range of planet masses we take, this approximate point of destruction by Joule heating occurs well before the planet can reach its Roche limit (the end point of each curve). Compared to the present location and the estimated age of Kepler-1520 b \citep{Morton16}, its progenitor could be on the order $\sim$0.1--0.3$M_\oplus$ that originated at $\sim$1 day. Converting to radius assuming $R_p = R_\oplus (M_p/M_\oplus)^{1/4}$, we find 8 such planets inside a day in the NASA Exoplanet Archive, which is smaller than the estimated $\sim$10--100 progenitors per disintegrating planet by \citet{Perez-Becker13}, but greater than zero. More precise estimate of the progenitor occurrence rate requires better knowledge of the conductivity profile of terrestrial planets and convolving over the distribution of stellar magnetic field strength.

In our simple calculation above, we assumed all the Joule heating on the planet to be deposited into the deep interior of the planet.\footnote{Given the uncertainty in planet vs.~stellar resistance, we also considered the total circuit Joule heating to be deposited into the planet and found that it immediately exceeds the self-binding energy of the planet we consider. If instead this heat is fully radiated away assuming the planet to reach some equilibrium temperature, the nightside temperature reaches $\sim$200-700 K, corresponding to peak emission at $\sim$5--15$\mu$m, which may be detectable with full-orbit phase curves with the MIRI instrument on JWST.} If the heating only reaches the upper crust, planet destruction could be expedited with each layer being lifted off. In addition, the deposited Joule heat flux is comparable to and exceeds the tidal heat flux on Io $\sim$1--3 W m$^{-2}$ \citep{Veeder94,Rathbun04}, believed to drive a runaway melting in the interior and to sustain its extreme volcanism \citep[e.g.,][]{Peale79}. Small rocky planets at the edge of the magnetic drag may also undergo significant melting and active surface volcanism, which could be detectable by the James Webb Space Telescope \citep[e.g.,][]{Seligman24}. A more careful calculation of thermal evolution would be necessary taking into account the partitioning of the heat into each layers of the planetary interior, internal phase transition, and heat transport through the melt-solid mixture \citep[see e.g.,][for recent attempts in this avenue]{Curry24,Herath24}. 

Another signature of star-planet magnetic interaction is nonthermal radio emission which is emitted at the cyclotron frequency \citep{Zarka01,Treumann06}. For mature stars that are $\sim$5 Gyr, $B_\star \sim 10$G and the cyclotron frequency $\sim$27 MHz which is not high enough for current or upcoming ground-based radio instruments \citep{Callingham24}. We would have a better chance of detecting such a signature on younger stars hosting a close-in rocky planet that is undergoing magnetic drag. Alternatively, star-planet magnetic interaction may be probed by detecting modulation in activity signature in stellar chromosphere that correlates with the planet's orbital period \citep[e.g.,][]{Shkolnik03,Shkolnik13}. Observational searches thus far have focused on hot Jupiters as larger planets are expected to drive larger Joule heating but they are difficult to distinguish from tidal dissipation. For sub-Earth rocky planets, magnetic drag dominates over tidal decay so ultra-short period rocky planets may provide a cleaner probe of star-planet magnetic interaction, in spite of their weaker Joule heating.

\subsection{Caveats and Suggestions}

Our analytic model, while describing remarkably well the observed edges in rocky planet population, is predicated on a few caveats, some of which are carried forward into our numerical simulations. We close our paper with a description of the major uncertainties, cautions for future modeling, and how these uncertainties may be constrained with future observations.

First, the outcome of photoevaporative mass loss is sensitive to the assumed evolution of stellar XUV luminosity. While directly constraining the full X-ray and EUV spectrum of individual stars is nigh impossible, analysis of solar data suggests that the EUV radiation decays more gradually than X-ray and dominates well beyond $\sim$100 Myr with the cumulative irradiation energy more heavily weighted toward later time \citep[e.g.,][]{King21}. The gradual decline of XUV luminosity suggests that caution must be taken when using the age of the system to determine the dominant channel of mass loss---the steep decline of hot sub-Neptunes and/or the increase in super-Earth/sub-Neptune fraction beyond 1 Gyr \citep[e.g.,][]{Berger20, Christiansen23} is not necessarily evidence against photoevaporation.

Second, the tidal decay depends on the assumed tidal quality parameter $Q'_\star$. While its orbital period dependence has been derived from hot Jupiter data \citep{Penev18}, its dependence on stellar type remains uncertain, especially for high ($>1.1 M_\odot$) and low ($<0.6 M_\odot$) mass stars. Observationally, while recent occurrence rate studies around A-type stars report a reduced population of sub-Neptunes and sub-Saturns around A stars compared to FG dwarfs, whether the same is true for super-Earths remains unclear \citep[see][their Figure 11]{Giacalone25}. If tidal dissipation is inefficient in A stars (from their lack of outer convective shell), we would expect larger population of short-period rocky planets so long as they form around A stars just as frequently as lower mass stars \citep[but see, e.g.,][]{Yang20,Chachan23}. Increasing the number of confirmed planets around these stars and measuring with precision the planet mass (see Appendix \ref{app:st-bin} for further discussion on the shortcomings in this avenue for the currently available data) and the system age will be crucial to constrain $Q'_\star$ by comparing where these planets lie in the mass-period diagram against our equation \ref{eq:Mp-tide-Q}.

Finally, the efficacy of the magnetic drag relies on the assumed time evolution of the stellar magnetic field strength and morphology. Our model describes well the observed edge in the radius-period space of small rocky planets within the currently reported range of stellar magnetic field intensity at $\sim$100 Myr \citep{Vidotto14} although the reality may be more complicated than our assumption of the dipolar field \citep[e.g.,][]{Garraffo18,Metcalfe21}. We may also expect generally stronger magnetic field for fully convective stars and weaker magnetic field for A-type stars which could translate to fewer (more) ultra-short sub-Earths around low-mass (high-mass) stars. Current data show the opposite (see Figure \ref{fig:RvP-rocky-alfven}) because the Joule heating rate scales with $R_\star^3$. To see the effect of weaker magnetic field for higher mass stars under our hypothesis (equation \ref{eq:P_mag_drag_scl}), the $B_\star$-$M_\star$ scaling needs to be steeper than $M_\star^{-7/3}$. More complete survey of planets with $<R_\oplus$ and $<1$ day period with {\it Transiting Exoplanet Survey Satellite (TESS)} and {\it Plato} over a wide range of stellar mass could help constrain $B_\star$-$M_\star$ scaling. Concomitantly, better knowledge of stellar magnetic fields over age and stellar type through e.g., Zeeman-Doppler Imaging \citep[e.g.,][]{Klein21} for stars hosting short-period rocky planets would be welcome to more definitively test the magnetic drag model as the origin of ultra-short period small planets and the catastrophically evaporating planets.

\vspace{0.3cm}
This project was conceived at the Density Matters program at the Ringberg Castle in Winter 2024---we thank the organizers of that meeting.
We thank the anonymous referee for an insightful report that helped improve our manuscript. We also thank Nicolas Cowan and Norm Murray for helpful discussions.
E.J.L. gratefully acknowledges support by NSERC, by FRQNT, by the Trottier Space Institute, and by the William Dawson Scholarship from McGill University. J.E.O. is supported by a Royal Society University Research Fellowship. This project has received funding from the European Research Council (ERC) under the European Union’s Horizon 2020 research and innovation programme (Grant agreement No. 853022).
This research has made use of the NASA Exoplanet Archive, which is operated by the California Institute of Technology, under contract with the National Aeronautics and Space Administration under the Exoplanet Exploration Program. 

\facility {Exoplanet Archive}

\software{We used the following python packages: scipy \citep{2020SciPy-NMeth}, numpy \citep{harris2020array}, matplotlib \citep{Hunter:2007}, pandas \citep{mckinney-proc-scipy-2010}.}

\appendix 
\counterwithin{figure}{section}
\counterwithin{equation}{section}

\section{Choice of Stellar Mass Bins}
\label{app:st-bin}

\begin{figure*}
    \centering
    \includegraphics[width=\textwidth]{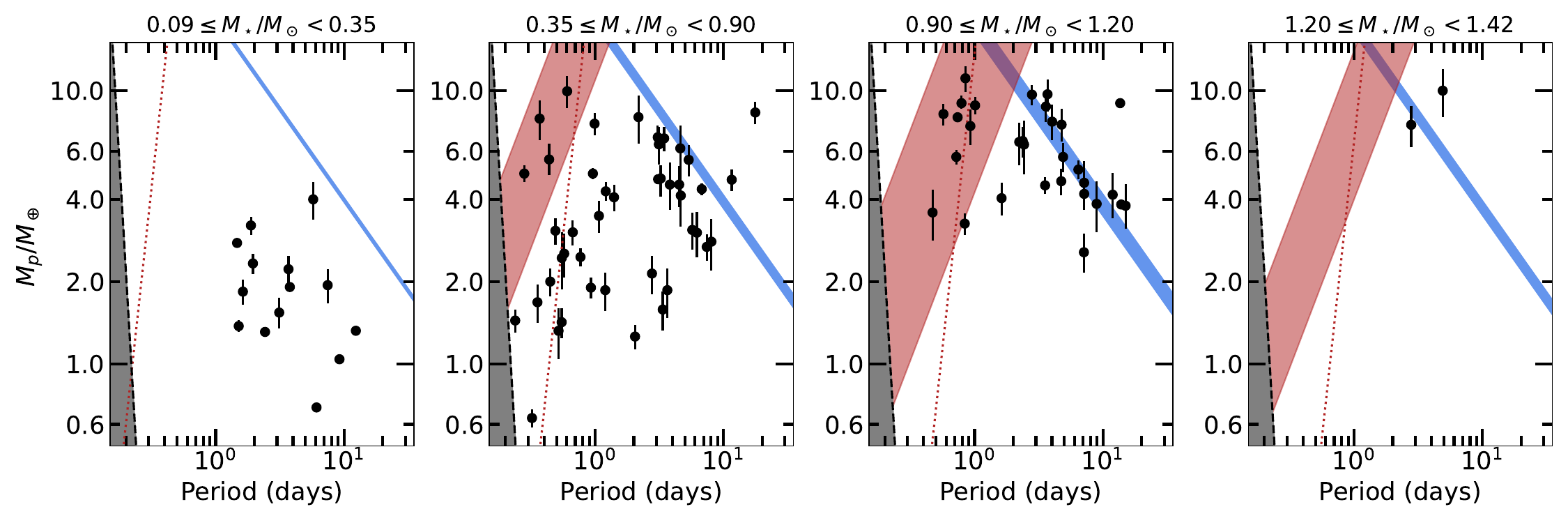}
    \caption{Same as Figure \ref{fig:Mp_Per_Mstar} but with the upper limit of the lowest stellar mass bin adjusted to 0.35 $M_\odot$ and the lower limit of the highest stellar mass bin adjusted to 1.2$M_\odot$.}
    \label{fig:Mp-Per-Mstar-diff}
\end{figure*}

Tidal dissipation inside the star is expected to occur most efficiently in the convective zone. We may therefore expect the edges in the small planet population carved by star-planet tidal interaction show noticeable change across the mass below/above which the stellar structure changes. Structural evolution models report stars below $\sim$0.35$M_\odot$ to become fully convective \citep[e.g.,][]{Chabrier97}. For these low mass stars, we may expect stronger level of tidal dissipation and therefore smaller $Q'_\star$. On the other end of the spectrum, hot stars above the Kraft break (effective temperature $\gtrsim$6100 K, $M_\star \geq 1.2 M_\odot$) are found to have a convective core and a radiative envelope and are therefore expected to have reduced tidal damping \citep[e.g.,][]{Winn10}. In Figure \ref{fig:Mp-Per-Mstar-diff}, we replot the mass-period diagram of rocky planets, adjusting the lowest and the highest stellar mass bins to identify any changes across 0.35$M_\odot$ and 1.2$M_\odot$ boundaries.

Around stars with mass $\geq 1.2M_\odot$, there are so few planets with mass measurements that we are unable to test our theories. The lack of data here is expected. First, both the planet-occurrence rate and the fraction of stars harboring sub-Neptunes are significantly reduced for F-type stars compared to G-type stars \citep[e.g.,][]{Mulders15,Yang20,Kunimoto20}. Second, there are fewer stellar spectral lines available for radial velocity follow-up for more massive stars, making mass measurements challenging, and the rapid rotation of massive stars can broaden the lines, further exacerbating the challenge.

Around stars with mass $<0.35 M_\odot$, we see two significant differences in the population of planets with precise mass measurements compared to those around more massive stars. First, the shortest orbital period is $\sim$1 day. At present, we cannot determine with certainty whether this is real or an observational bias introduced by e.g., survey design (e.g., {\it Kepler} focused on FGK type stars, and the input stellar catalog for TESS is more sensitive to stars hotter than $\sim$4000 K, \citealt{Stassun19}). Given the rarity of ultra-short period planets \citep{SanchisOjeda2014}, we may not have large enough baseline dataset to detect these planets around low mass stars.
If, on the other hand, the lack of ultra-short period planets is real around stars $<0.35 M_\odot$, the theory of tidal inspiral implies $Q'_\star \sim 10^4$ for both fixed $Q'_\star$ and for the normalization of period-varying $Q'_\star$. Curiously, \citet{Wu24} predict a dearth of ultra-short period small planets around low mass, fully-convective M dwarfs using the theory of resonance locking \citep[e.g.,][]{Ma21}. However, inspiral under complete resonance locking would predict no mass-dependence so we may expect a vertical edge in the mass-period plot, and it will be difficult to explain the maximum mass of 10$M_\oplus$ for rocky planets across a range of stellar mass. If the planets break out of resonance with the internal stellar modes, we may recover the planet mass-dependent inspiral time \citep{Wu24}.

Yet another difference we see in the current rocky planet population around lowest mass stars is that the maximum mass is $\sim$4$M_\oplus$ instead of 10$M_\oplus$ (we verify the same decline in the maximum mass in this stellar mass bin with the sample studied by \citealt{Parc24}). While radial velocity measurements of such cool stars are challenging due to the complex spectra and higher activity signature of cooler dwarfs, if current instruments are sensitive to $\leq 4 M_\oplus$ objects, they would also be sensitive to higher mass objects. The overall lower masses of rocky planets around these cool stars may therefore be real. One reason for the reduced mass may be the lack of initial dust mass reservoir. \citet{Chachan23} find the expected fraction of stars with inner super-Earths to drop for stars $\lesssim$0.4--0.6$M_\odot$ because their initial disk mass are too small. We experimented with the lowest stellar mass bin and found that the upper mass limit to rocky planets stays $\sim$4$M_\oplus$ out to $M_\star < 0.45 M_\odot$, consistent with the expectation of \citet{Chachan23}.

\section{Integral of the Alfv\'{e}n Drag}
The analytic form of the indefinite integral on the left hand side of equation \ref{eq:integral-alfven} is
\begin{equation}
    \int \frac{x^3}{X} dx = 2A^4 X \times \left[\frac{1}{384}\left(\frac{x}{A}\right)^4 (-105 X^3 + 385 X^2 - 511 X + 279) + \frac{35 \tanh(X)}{128 X}\right]
    \label{eq:integral-analytic}
\end{equation}
where $x \equiv a/a_0$, $X \equiv (1-A/x)^{1/2}$, $A \equiv R_\star/a_0$.

\bibliography{max-core}{}
\bibliographystyle{aasjournal}

\end{document}